\documentclass[aps,prl,twocolumn,showpacs]{revtex4}
\usepackage{amssymb}
\usepackage{graphicx,amsmath}

\newcommand{\bea}{\begin{eqnarray}}
\newcommand{\eea}{\end{eqnarray}}
\newcommand{\beq}{\begin{equation}}
\newcommand{\eeq}{\end{equation}}
\newcommand{\benu}{\begin{enumerate}}
\newcommand{\enu}{\end{enumerate}}

\newcommand{\al}{\alpha}
\newcommand{\be}{\beta}

\newcommand{\om}{\omega}
\newcommand{\Om}{\Omega}
\newcommand{\ep}{\epsilon}

\newcommand{\lam}{\lambda}

\newcommand{\ham}{\mathcal{H}}
\newcommand{\ord}{\mathcal{O}}

\newcommand{\ptl}{\partial}

\newcommand{\bk}{{\bf k}}
\newcommand{\bq}{{\bf q}}

\newcommand{\br}{{\bf r}}
\newcommand{\bu}{{\bf u}}
\newcommand{\bM}{{\bf M}}

\begin{document}

\title{
Magneto-elastic quantum fluctuations and phase transitions
in the iron superconductors
}
\date{\today}
\author{I. Paul}
\affiliation{
Institut N\'{e}el, CNRS/UJF, 25 avenue des Martyrs, BP 166,
38042 Grenoble, France
}

\begin{abstract}
We examine the relevance of magneto-elastic coupling to describe the
complex magnetic and structural behaviour of the different classes of the
iron superconductors. We model the system as a two-dimensional
metal whose magnetic excitations interact with the distortions of the
underlying square lattice. Going beyond mean field we find that
quantum fluctuation effects can explain two unusual features of these
materials that have attracted considerable attention.
First, why iron telluride
orders magnetically at a non-nesting wave-vector $(\pi/2, \pi/2)$
and not at the nesting wave-vector $(\pi, 0)$ as in the iron arsenides,
even though the nominal band structures of both these systems are similar.
And second, why the $(\pi, 0)$ magnetic transition in the iron arsenides
is often preceded by an orthorhombic structural transition. These are robust
properties of the model, independent of microscopic details, and they emphasize
the importance of the magneto-elastic interaction.

\end{abstract}

\maketitle
\emph{Introduction.}--- The recently discovered iron
superconductors with unusually high transition temperatures exhibit
a rich phase diagram that includes structural, magnetic and
superconducting transitions~\cite{kamihara,review}.
As such these materials are the latest
playgrounds to study how in complex materials different phases
compete, and how this unconventional setting eventually gives rise
to superconductivity. Theoretically, one of the
goals at present is to identify the microscopic interactions that
give rise to the rich phase diagram. This motivates us to present
a microscopic study of the simplest symmetry-allowed model Hamiltonian
describing magneto-elastic interaction. The results allow us to
argue that this coupling contains physics relevant for
the iron superconductors, and therefore
it is an important microscopic ingredient.

Crystallographically these materials have a layered structure, which
is reflected in their energy bands with weak dispersion along the $c$-axis
compared to that along the $ab$-plane~\cite{singh,subedi}. Consequently,
it is often simpler to consider them as two-dimensional systems weakly coupled
along the $c$-axis.
The undoped and the lightly doped compounds usually undergo magneto-structural
transitions from
paramagnetic metals with tetragonal crystalline symmetry to low temperature
antiferromagnetic (AF) metals with either orthorhombic (in case of the FeAs systems)
or monoclinic (in case of Fe$_{1+y}$Te) structures.
These transitions are suppressed in favour of superconductivity when they
are either doped or put under external pressure.
The AF order of the FeAs systems is at the wave-vector
$(\pi,0)$ in the Brillouin zone defined by the plane of the Fe atoms with 1Fe/cell,
and it is often preceded in temperature by the structural transition.
On the other hand, the AF order of Fe$_{1+y}$Te is at
$(\pi/2, \pi/2)$, and the lattice distorts simultaneously.

While it is likely that the $(\pi,0)$
order of FeAs is a consequence of the Fermi surface nesting in these multi-band
systems~\cite{singh}, from the perspective of a band picture there
remains at least two important puzzles concerning the magneto-structural properties.
First, why Fe$_{1+y}$Te, whose nominal band structure is similar to
that of the FeAs systems~\cite{subedi},
orders at $(\pi/2, \pi/2)$ and not at the nesting wave-vector $(\pi,0)$. And second,
why the AF transition of the FeAs systems is often preceded in temperature
by a tetragonal-orthorhombic structural
transition. The main results of this study are to show that both these features
are natural consequences of quantum fluctuations induced by the
magneto-elastic interaction. We find that the former property is due to
spin fluctuations scattering with short wavelength phonons, and the latter is
driven by critical spin fluctuations near a $(\pi, 0)$ AF transition.

The above questions, as well as the general magneto-structural properties of these materials,
have been addressed earlier from various points of view. Some of these are based on
itinerant electron models which emphasize the physics
of nested Fermi surfaces~\cite{zhang-knolle,han}.
There are also studies that suggest that the electron-electron
interaction is strong~\cite{haule-aichhorn},
which justifies describing the magnetic properties by Heisenberg
spin models with appropriate couplings~\cite{qimiao-fang}.
The structural transition has been viewed as a consequence of various
kinds of electron order such as orbital ordering~\cite{orbital},
spin nematic ordering~\cite{henley-chandra} at temperatures above the
magnetic transition~\cite{fang,fernandes}, as well as the ordering of
orbital currents~\cite{kang}.

In the past there has been few studies of the magneto-elastic properties of the
FeAs materials which concentrated on the $c$-axis motion of the As atoms, and its
strong influence on the Fe-As bond and eventually on the magnetism~\cite{past}.
These are motivated by the observation of a pressure-driven first order volume collapse
transition in CaFe$_2$As$_2$, which is concomitant with the loss of magnetism~\cite{kreyssig}.
On the other hand, a microscopic study of the influence of the $ab$-plane distortions
of the lattice on the magnetic sector using an effective model is currently lacking.
This is despite the fact that the in-plane distortions have an important symmetry-allowed
coupling with the order parameters of the magnetic transition,
which can make the transition weakly first
order~\cite{barzykin,larkinpikin}.

Here we perform such a microscopic study using a model Hamiltonian $\ham_{ME}$
introduced phenomenologically in Ref.~\onlinecite{paul}, and where it was argued
that a mean field treatment of $\ham_{ME}$ describes several
salient features of the magneto-structural transitions in the FeAs~\cite{cano}
and Fe$_{1+y}$Te~\cite{paul} systems.
In the case of FeAs it
explains (i) why the magnetic and structural transitions are separate in some materials while
they are concomitant in others, (ii) why the transitions appear to be first order in some cases
even though they are allowed to be second order from a symmetry point of view, and
(iii) why the systems prefer a collinear magnetic state instead of a non-collinear
order~\cite{cano}.
In the case of Fe$_{1+y}$Te it explains why in the magnetic phase the system undergoes
a uniform monoclinic distortion, and a modulated one with wave-vector $(\pi,\pi)$
such that the Fe-Fe bonds are alternately elongated and shortened~\cite{paul}.
Our aim here is to go beyond mean field, and to study the quantum fluctuations of
the elastic and the magnetic degrees of freedom and thereby examine further the relevance
of $\ham_{ME}$.
\begin{figure}[!!t]
\begin{center}
\includegraphics[width=8cm]{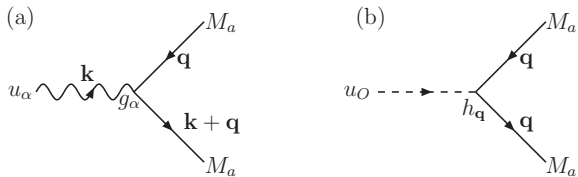}
\caption{
Magneto-elastic coupling [Eq.~(\ref{eq:hamME1})]
describing scattering of paramagnons (straight line) (a) with phonons
(wavy line) at finite $\bk$, and (b) with orthorhombic distortion
(dash line) at $\bk \rightarrow 0$ limit. The associated
matrix elements $g_{\al}(\bk,\bq)$ and $h(\bq)$ are defined in
Eqs.~(\ref{eq:g}) and (\ref{eq:h}) respectively.
}
\label{fig1}
\end{center}
\end{figure}

\emph{Model.}--- We consider a two-dimensional metal on a square
lattice with magnetic and elastic degrees of freedom that are
coupled by the Hamiltonian \beq \label{eq:hamME1} \ham_{ME} = \lam_0
\sum_{\langle i j \rangle} \left( \bu_i - \bu_j \right) \cdot
\hat{n}_{ij} \left( \bM_i \cdot \bM_j \right). \eeq Here $\lam_0$ is
the coupling constant with dimension of energy per length, $\bu_i
\equiv \bu(\br_i)$ is the displacement of the Fe atom from its
equilibrium position at $\br_i$, $\langle ij \rangle$ implies
nearest neighbour sites, $\hat{n}_{ij}$ is the unit vector along the
$i-j$ bond, and $\bM_i$ is the local magnetization at $\br_i$. We
neglect spin-orbit coupling, and magneto-elastic terms of
$\ord (u^2)$ and higher.
In the context of insulating magnets $\ham_{ME}$ describes the
variation $\ptl J/ \ptl r$ of the Heisenberg exchange $J$ with bond
length, and its effects are well-studied~\cite{becca-penc}. In the
case of metals, where such couplings are far less studied,
$\ham_{ME}$ describes the bond-length dependence of the parameter
$V_{ij}$ that characterize the nearest neighbour interaction
$V_{ij}\rho_{i \uparrow} \rho_{j \downarrow}$, with  $\rho_{i
\sigma}$ being the density of electrons with spin $\sigma$ at site
$i$.

We study the system from the paramagnetic side, and we approximate
$\bM_i$ to be $O(3)$ variables describing spin fluctuations (paramagnons).
Conceptually, they can be introduced as Hubbard-Stratonovich fields to decouple the
appropriate electron-electron interaction, after which the electrons can be integrated
out~\cite{moriya}. The resulting action for the magnetic sector can be written as
$
S_M = \sum_{\bq, \nu_n, a} \chi_0^{-1} (\bq, \nu_n) M^{\ast}_{a} (\bq, \nu_n) M_{a} (\bq, \nu_n),
$
where $a = (x, y, z)$ in $O(3)$ space, $\bM(\bq)$ is the Fourier transform of $\bM(\br_i)$,
and $\nu_n$ is a bosonic Matsubara frequency. We assume that the paramagnon
propagator describes damped dynamics with
$
\chi_0^{-1}(\bq, \nu_n) = \Pi_0 (\bq) + \left| \nu_n \right|,
$
and having the bare dispersion
$
\Pi_0 (\bq) = \Om_0 + \Om_1 [ \cos q_x \cos q_y - (\cos 2q_x + \cos 2q_y)/8]
$
with $\Om_0 > \Om_1 > 0$.
With this choice $\Pi_0 (\bq)$ has a
global minima at $\bq = (\pi, 0)$ (and at the symmetry related points) which models
the nesting property of the underlying Fermi surface.
The approach towards the magnetic instability can be described by lowering $\Om_0$.

As our results are independent of the microscopic details of the elastic sector,
it is sufficient to describe it using the simplest model compatible with the
square symmetry.
The lattice variables are defined by the strain tensor
$
\ep_{\al \be} (\br_i) = u_{\al \be} + i/2 \sum_{\bk \neq 0} [ k_{\al} u_{\be}(\bk)
+ \al \leftrightarrow \be ] \exp(i \bk \cdot \br_i),
$
where $\al, \be = (x,y)$ in two dimensions. This includes the uniform strains $u_{\al \be}$,
and the phonons described by $\bu(\bk)$ which is
the Fourier transform of $\bu(\br_i)$~\cite{larkinpikin}.
The energy per unit area due to $u_{\al \be}$ is
given by
$
E_S = C_{11} (u_{xx} + u_{yy})^2/2 + C_{12} u_{xx}u_{yy} + C_{66} u_{xy}^2/2,
$
where the $C$s are the elastic constants in Voigt notation.
The order parameter for the tetragonal-orthorhombic transition
is $u_o \equiv u_{xx} - u_{yy}$, and the associated elastic
constant is $C_o \equiv (C_{11} - C_{12})/2$.
Next, the phonons are described by the action
$
S_P = \sum_{\bk, \nu_n, \mu} D_{\mu}^{-1} (\bk, \nu_n) u^{\ast}_{\mu} (\bk, \nu_n)
u_{\mu} (\bk, \nu_n),
$
where $\mu$ is the polarization index. We assume the propagator to have the standard form
$D_{\mu}^{-1}(\bk, \nu_n) = - 2 \om_{\bk \mu}/(\nu_n^2 + \om^2_{\bk \mu})$,
with $\om_{\bk \mu}$ the phonon dispersion.
We obtain it from a harmonic theory in which the eigenvalues of
the dynamical matrix $N_{\al \be}(\bk)$ give $m \om^2_{\bk \mu}$, $m$ being the atomic mass
of Fe. We take $N_{xx} = 4C_{11} \sin^2 (k_x/2) + C_{66} \sin^2(k_y/2)$,
$N_{yy} = N_{xx} (x \leftrightarrow y)$, and
$N_{xy} = N_{yx} = (4C_{12} + C_{66}) \sin(k_x/2) \sin(k_y/2)$.
Thus, the elastic sector is entirely characterized by the $C$s and $m$.

Before performing the calculations
it is instructive to re-write the interaction
in Fourier space as
\beq
\ham_{ME} = 2i \lam_0 \sum_{\al, a, \bk, \bq} g_{\al} (\bk, \bq) u_{\al} (\bk)
M^{\ast}_{a} (\bk+\bq) M_{a} (\bq),
\nonumber
\eeq
where
$g_{\al}(\bk, \bq)$ are matrix elements with
\beq
\label{eq:g}
g_{\al} (\bk, \bq) = \sin (k_{\al} + q_{\al}) - \sin (q_{\al}).
\eeq
They play an important role due to the property that $g_{\al} \rightarrow 0$ as $k \rightarrow 0$,
which implies that the coupling to the phonons vanishes in the limit of uniform displacements.
In this limit the paramagnons couple to the classical variables $u_{\al \be}$, and in particular
the coupling to $u_o$
is associated with the matrix element
\beq
\label{eq:h}
h(\bq) = \cos q_x - \cos q_y.
\eeq
The coupling of the paramagnons to the phonons ($\bk \neq 0$) and to $u_o$ ($\bk \rightarrow 0$)
along with their respective matrix elements are shown graphically in Fig.~\ref{fig1}.

As our model is two dimensional, strictly speaking it cannot be used to
study the finite temperature magnetic transition due to the Mermin-Wagner theorem.
In the following our strategy is to
perform calculations at zero temperature ($T$) where the theorem is inapplicable,
and to infer finite-$T$ consequences using adiabaticity argument. In general, at finite-$T$
the magnitude of the results below are enhanced due to the thermal fluctuations.
\begin{figure}[!!t]
\begin{center}
\includegraphics[width=8cm]{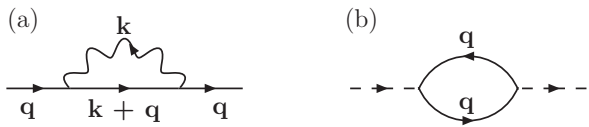}
\caption{Magneto-elastic coupling induced quantum fluctuations
producing
(a) static paramagnon self-energy $\Pi_2(\bq)$ due to
scattering with phonons, and (b) correction to the orthorhombic
elastic constant $C_o$ due to paramagnons.
}
\label{fig2}
\end{center}
\end{figure}
\begin{figure}[!!t]
\begin{center}
\includegraphics[width=8cm,height=6.5cm]{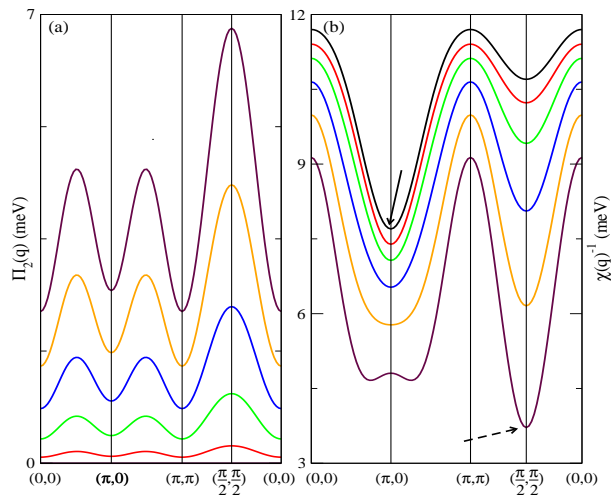}
\caption{
(colour online). (a) Paramagnon self-energy $\Pi_2(\bq)$ along the high symmetry directions
of the Brillouin zone for increasing values of magneto-elastic coupling
$\lam_0 = 1, \cdots, 5$ (bottom to top) in unit of 10 meV/\AA. It is always peaked at
$(\pi/2,\pi/2)$. (b) The corresponding renormalized paramagnon dispersions
$\chi^{-1} (\bq) = \Pi_0 (\bq) - \Pi_2(\bq)$
(second-from-top to bottom curves respectively),
and the bare dispersion $\Pi_0(\bq)$ (topmost curve, shifted up by 0.2 meV for clarity).
At large enough $\lam_0$ (bottom-most) the global minima changes to $(\pi/2,\pi/2)$
(dash arrow)
from the nesting driven $(\pi,0)$ wave-vector (solid arrow).
}
\label{fig3}
\end{center}
\end{figure}

\emph{Results.}---
(i) First we examine the static paramagnon self-energy obtained due to virtual scattering with the
phonons. This is given by (see Fig.~\ref{fig2}a)
\bea
\Pi_2 (\bq) &=& - 4 \lam_0^2 T \sum_{\bk, \nu_n, \al, \be, \mu} (2m \om_{\bk \mu})^{-1}
g_{\al} (\bk, \bq) g_{\be} (\bk, \bq)
\nonumber \\
&\times&
\ep_{\mu}^{\al}(\bk) \ep_{\mu}^{\be}(\bk)
D_{\mu} (\bk, i \nu_n) \chi_0 (\bk + \bq, i \nu_n),
\nonumber
\eea
where $\hat{\ep}_{\mu}(\bk)$ are the polarization vectors.
We perform the $\bk$-sum numerically, taking
$(\Om_0, \Om_1) = (10, 2)$ meV (the precise value of these parameters do not affect
the result qualitatively).
We also use
$(C_{11}, C_{12}, C_{66}) = (0.484, 0.161, 0.141)$ $10^4$ meV/\AA$^2$
from elastic constant measurements on BaFe$_2$As$_2$~\cite{yoshizawa}
suitably normalized for two dimensions.
In Fig.~\ref{fig3}a we plot $\Pi_2(\bq)$
along high symmetry directions of the Brillouin zone for
various values of the coupling $\lam_0$. Note that the magnitude of
$\Pi_2(\bq)$ increases with $\lam_0$ as expected, but more importantly,
it is peaked at $(\pi/2, \pi/2)$ for all $\lam_0$
(below we explain why). The consequence of this property is demonstrated
in Fig.~\ref{fig3}b where we show the corresponding renormalized
paramagnon dispersions (second-from-top to bottom curves) given by
$
\chi^{-1}(\bq) = \Pi_0 (\bq) - \Pi_2 (\bq),
$
as well as the bare dispersion $\Pi_0(\bq)$ for comparison (topmost curve).
Since $\Pi_2(\bq)$ is always peaked at $(\pi/2, \pi/2)$,
for $\lam_0 > \lam_0^{\ast}$ (e.g., the bottommost curve
in Fig.~\ref{fig3}b) the global minima of
$\chi^{-1}(\bq)$ changes from the nesting driven wave-vector $(\pi,0)$ to the magneto-elastic
coupling driven wave-vector $(\pi/2, \pi/2)$. This implies that in this model
the magnetic instability is at $(\pi/2, \pi/2)$ for sufficiently large $\lam_0$.
Thus, the $(\pi/2, \pi/2)$ magnetic order observed in Fe$_{1+y}$Te
can be explained as a signature of strong magneto-elastic
coupling, while in the FeAs systems this effect is presumably weaker and the nesting driven
$(\pi, 0)$ instability is preferred. Indeed, several authors have argued that
the electron-electron interaction effects, which favours the magneto-elastic
coupling, is stronger in Fe$_{1+y}$Te than in the FeAs systems~\cite{review,orbital}.

The above result can be understood simply from the following argument. Since $g_{\al}$
vanishes for $k \rightarrow 0$, the $\bk$-sum above is dominated by large $k$ (short
wavelength phonons), and
typically $\om_{\bk \mu} \sim E_0$ the phonon bandwidth, which is the largest
energy scale in the model. Thus,
$
T \sum_{\nu_n} D_{\mu} (\bk, i \nu_n) \chi_0 (\bk + \bq, i \nu_n) \sim 1/E_0,
$
and its $\bq$-dependence is negligible. Therefore, the dominant $\bq$-dependence of
$\Pi_2 (\bq)$  is from the matrix elements $g_{\al} g_{\be}$. Then a simple power counting
argument, and the fact that in a stable lattice the diagonal components
of $N_{\al \be}$ are larger than the off-diagonal ones, establishes that
\beq
\label{eq:Pi2}
\Pi_2(\bq) = A + B \left( \sin^2 q_x + \sin^2 q_y \right) - C \cos q_x \cos q_y,
\eeq
with $A, B > C > 0$.
Note that this result is independent of the details of the paramagnon and the phonon
dispersions, and its main ingredient is the fact that $\ham_{ME}$
involves the nearest neighbour sites. Including couplings with longer range
will not change the result provided they are smaller in magnitude compared to $\lam_0$.
This is plausible, since the electron-electron interaction is expected to reduce
with distance. However, it needs to be checked using first principles calculation.
Note also that the value of $\lam_0^{\ast}$
depends intricately on the microscopic
details. In particular, it depends on the relative strength of the nesting driven minimum at
$(\pi, 0)$, and on the magnitude of the longer range magneto-elastic couplings.

(ii) Next we compute the correction to the orthorhombic elastic constant $C_o$ due to
quantum fluctuations of the paramagnons which is given by (see Fig.~\ref{fig2}b)
\beq
\delta C_o = - 3 \lam_0^2 T \sum_{\bq, \nu_n} h^2(\bq) \chi_0^2 (\bq, i\nu_n).
\nonumber
\eeq
For two dimensional spin fluctuations in the vicinity of
a $(\pi,0)$ magnetic instability (which is tuned by $\Om_0$) we get
\beq
\label{eq:dC}
\delta C_o = - 4 \lam_0^2/(\pi^2 \Om_1) \log (1/\delta),
\eeq
where $\delta = (\Om_0 - \Om_{\rm cr})/\Om_1$ measures the closeness
to the magnetic instability at $\Om_0 = \Om_{\rm cr}$.
 This implies that, before the system becomes magnetic
($\Om_0 > \Om_{\rm cr}$), the renormalized elastic constant $(C_o + \delta C_o)$
vanishes and the system encounters an orthorhombic instability. This softening
of the lattice with $\delta C_0 \propto \log (1/\chi_m)$, where $\chi_m$ is the dimensionless
static magnetic susceptibility at $(\pi, 0)$, is in principle verifiable from
neutron scattering data and from elastic constant measurements.
A similar softening of $C_0$ due to nematic fluctuations is discussed in
Ref.~\onlinecite{fernandes}.
Using adiabaticity
argument it is possible to extend the phase boundaries to finite temperatures in
the $T-\Om_0$ plane, and the resulting phase diagram shows that the magnetic transition
is preceded in temperature by the orthorhombic instability. However, close enough to the
finite-$T$ magnetic transition the critical theory is non-Gaussian, and the
quantitative aspects of the current treatment becomes invalid.

In practice, even at $T=0$ and in
three dimensions the log divergence is cutoff by the paramagnon dispersion along
the $c$-axis, and the precise fate of the structural instability depends on microscopic
details. Nevertheless, for quasi-two dimensional paramagnons one should expect
considerable orthorhombic softening, and therefore this mechanism explains
why often (but not always) the $(\pi,0)$ magnetic
transition in the FeAs systems is accompanied by an orthorhombic transition. In passing we
note that a similar monoclinic softening is expected near a $(\pi/2, \pi/2)$ magnetic
transition (and can be relevant for Fe$_{1+y}$Te), but to capture this physics one needs
to generalize $\ham_{ME}$ and include next-nearest neighbour couplings.

\emph{Conclusion.}---
We studied the effects of quantum fluctuations induced by magneto-elastic coupling
[Eq.~(\ref{eq:hamME1})] in a two-dimensional metal on a
square lattice. The coupling describes the simplest symmetry-allowed
scattering between the paramagnons and the distortions of the
lattice. At a qualitative level the model explains (i) why Fe$_{1+y}$Te
orders magnetically at $(\pi/2,\pi/2)$ and not at the nesting
wave-vector $(\pi, 0)$, and (ii) why the FeAs systems often undergo an
orthorhombic transition in the vicinity of the $(\pi, 0)$ magnetic transition.
The former result is due to paramagnons scattering with short wavelength phonons,
and the latter is driven by the critical spin fluctuations.
We hope these results will stimulate further studies of the coupling
using a more realistic model and in conjunction with
first principles calculations. Such couplings can give rise to qualitatively new
physics, and they can be relevant for correlated metals in general.

The author is very thankful to E. Boulat, A. Cano, H. Capellmann, E. Kats,
I. Vekhter,
and T. Ziman for insightful discussions.

\end{document}